# A Note On Steady Flow of Incompressible Fluid Between Two Co-rotating Disks

**Milan Batista**

University of Ljubljana, Faculty of Maritime Studies and Transportation

Pot pomorscakov 4, 6320 Portoroz, Slovenia, EU

milan.batista@fpp.edu

**Abstract**

The article provides an analytical solution of the Navier-Stokes equations for the case of the steady flow of an incompressible fluid between two uniformly co-rotating disks. The solution is derived from the asymptotical evolution of unknown components of velocity and pressure in a radial direction--in contrast to the Briter-Pohlhausen analytical solution, which is supported by simplified Navier-Stokes equations. The obtained infinite system of ordinary differential equations forms recurrent relations from which unknown functions can be calculated successively. The first and second approximations of solution are solved analytically and the third and fourth approximations of solutions are solved numerically. The numerical example demonstrates agreements with results obtained by other authors using different methods.

**1. Introduction**

The object of investigation of this paper is the steady flow of viscous fluid between two parallel co-rotating disks where the fluid enters an inner cylinder in a radial direction and emerges at the outer cylinder (Figure 1). Note that the problem differs from the celebrated von Karman problem and its generalization since this problem investigates the swirling flow induced by the rotation of infinite disks (Rajagopal 1992).

It seems that the problem--in the context of application in the design of centrifugal pumps--was first studied in 1962 by M.Breiter in K.Pohlhausen (1962). From the linearized boundary layer approximations of Navier-Stokes equations they derived the analytical expressions for velocity components and pressure showing that the solution depends on kinematic viscosity, angular velocity and the distance between the disks. They also provide a numerical solution of the non-linearized equations using the finite difference method with constant inlet profile. This line of research was continued by W.Rice and coworkers, whose main goal was the prediction of the performance of a centrifugal pump/compressor. They used different methods to obtain velocity and pressure distribution of flow between two disks. Thus Rice (1963) studied the flow with equations derived by using hydraulic treatment of bulk flow; Boyd and Rice (1968) used the finite difference method to calculate velocity and pressure for various parabolically distributed inlet velocities; and Boyack and Rice (1971) used what they called the integral method, in which the velocity components are represented by a polynomial of the axial coordinate.

Another line of research of the so called laminar source-sink flow in a rotating cylindrical cavity originated with the analytical study by Hide (1968), who gave the approximate asymptotic expressions for velocity components using the boundary-layer technique. Numerically, by using the finite-difference method, the problem was solved by Bennetts and Jackson (1974). Owen et al (1985) used the integral-momentum of von Karman that extended Hide's linear approximation to the turbulent flow. Recently, the steady flow between rotating disks was included in the study by Crespo del Arco et al (1996) using a pseudo-spectral numerical method.

From the above brief review of literature it is clear that the problem is analytically and especially numerically well studied and the results of calculations are in agreement with experiments. However, all available analytical solutions are based on variants of approximation. In this article, an alternative analytical solution will be presented, which is based on the asymptotic expansion of unknown functions by a method similar to those of Savage (1964), who considered the stationary radial directed flow between steady disks. The article is organized as follows. After the basic equations are established, their dimensionless forms are provided. The equations are solved and the results are compared with other methods.

**2 Basic equations**

Consider the steady axisymmetrical isothermal flow of incompressible viscous fluid between two co-rotating disks in the absence of body force. The disks have inner radius $a$ and outer radius $b$. The distance between disks is $2h$. Both disks rotate in the same direction with constant angular velocity $\Omega$. For description of flow, the reference frame rotating with angular velocity $\Omega$ is used. In this frame, by using the cylindrical coordinate system with coordinates $r$ and $z$, the continuity equation and Navier-Stokes equations have the form (cf. Acheson 1990, Landau and Lifshitz 1987):

$$\frac{1}{r}\frac{\partial}{\partial r}(ru) + \frac{\partial w}{\partial z} = 0 \qquad (1)$$

$$u\frac{\partial u}{\partial r} + w\frac{\partial u}{\partial z} - \frac{v^2}{r} - 2\Omega v = -\frac{1}{\rho}\frac{\partial p}{\partial r} + \nu\left(\frac{\partial^2 u}{\partial r^2} + \frac{1}{r}\frac{\partial u}{\partial r} - \frac{u}{r^2} + \frac{\partial^2 u}{\partial z^2}\right)$$

$$u\frac{\partial v}{\partial r} + w\frac{\partial v}{\partial z} + \frac{uv}{r} + 2\Omega u = \nu\left(\frac{\partial^2 v}{\partial r^2} + \frac{1}{r}\frac{\partial v}{\partial r} - \frac{v}{r^2} + \frac{\partial^2 v}{\partial z^2}\right) \qquad (2)$$

$$u\frac{\partial w}{\partial r} + w\frac{\partial w}{\partial z} = -\frac{1}{\rho}\frac{\partial p}{\partial z} + \nu\left(\frac{\partial^2 w}{\partial r^2} + \frac{1}{r}\frac{\partial w}{\partial r} + \frac{\partial^2 w}{\partial z^2}\right)$$

where $u(r,z)$, $v(r,z)$, $w(r,z)$ are the components of relative velocity in radial, tangential and axial direction, $p(r,z)$ is the reduced pressure, $\rho$ is the density, and $\nu$ is the kinematic viscosity.

Equations (1) and (2) are to be solved on the domain $r \in [a,b]$ and $z \in [-h,h]$ subject to the following boundary conditions along the disks' plane

$$u(r,\pm h) = v(r,\pm h) = w(r,\pm h) = 0 \qquad (3)$$

The boundary condition in entrance and outer cross section requires prescribed velocity components as functions of coordinate *z*. Since the asymptotic series solution, which will be used, does not offer enough free parameters to satisfy this boundary condition, it is replaced by prescribing the volume flow rate *Q*. Therefore, at outer cross section $r = b$ one has the condition

$$Q = 2\pi b \int_{-h}^{h} u(b,z) dz \qquad (4)$$

Because (4) does not refer to the interval $[a,b]$ its limits become artificial. So in this context *b* will be used as the reference radius. Also, because the boundary condition at the inner and outer cross sections will not be precise, the solution will not cover the inner source region and outer sink layer (Owen et al 1985).

Once equations (1) and (2) are solved, the tangential velocity components in the inertial reference frame is obtained by adding velocity $\tilde{v} = \Omega r$ to relative tangential velocity *v* and the total pressure obtained by adding the pressure $\tilde{p} = \rho \Omega^2 \frac{r^2}{2}$ to the pressure *p*.

**3. Dimensionless form of equations**

Equations (1), (2), (3) and (4) are made dimensionless by setting (Ames 1965)

$$r^* \equiv \frac{r}{b} \qquad z^* \equiv \frac{z}{h} \in [-1,1]$$
$$u^* \equiv \frac{u}{U} \qquad v^* \equiv \frac{v}{V} \qquad w^* \equiv \frac{w}{W} \qquad p^* \equiv \frac{p}{P} \qquad (5)$$

where $U$, $V$, $W$ and $P$ are appropriate scales of $u$, $v$, $w$ and $p$ respectively. Substituting dimensionless variables (5) into equations (1) and (2) assuming that $u^*$ and $v^*$ and also $\dfrac{\partial u^*}{\partial r^*}$ and $\dfrac{\partial w^*}{\partial z^*}$ are of the same orders, the following scales are yielded

$$U = V = b\Omega \qquad W = \left(\frac{h}{b}\right)U \qquad P = \rho U^2 \qquad (6)$$

By using (5) and (6) equations (1) and (2) become

$$\frac{1}{r^*}\frac{\partial}{\partial r^*}\left(r^* u^*\right) + \frac{\partial w^*}{\partial z^*} = 0 \qquad (7)$$

$$u^*\frac{\partial u^*}{\partial r^*} + w^*\frac{\partial u^*}{\partial z^*} - \frac{v^{*2}}{r^*} - 2u^* = -\frac{\partial p^*}{\partial r^*} + \frac{1}{\lambda^2}\left[\varepsilon^2\left(\frac{\partial^2 u^*}{\partial r^{*2}} + \frac{1}{r^*}\frac{\partial u^*}{\partial r^*} - \frac{u^*}{r^{*2}}\right) + \frac{\partial^2 u^*}{\partial z^{*2}}\right]$$

$$u^*\frac{\partial v^*}{\partial r^*} + \frac{u^* v^*}{r^*} + w^*\frac{\partial v^*}{\partial z^*} + 2v^* = \frac{1}{\lambda^2}\left[\varepsilon^2\left(\frac{\partial^2 v^*}{\partial r^{*2}} + \frac{1}{r^*}\frac{\partial v^*}{\partial r^*} - \frac{v^*}{r^{*2}}\right) + \frac{\partial^2 v^*}{\partial z^{*2}}\right] \qquad (8)$$

$$\varepsilon^2\left(u^*\frac{\partial w^*}{\partial r^*} + w^*\frac{\partial w^*}{\partial z^*}\right) = -\frac{\partial p^*}{\partial z^*} + \varepsilon^2\frac{1}{\lambda^2}\left[\varepsilon^2\left(\frac{\partial^2 w^*}{\partial r^{*2}} + \frac{1}{r^*}\frac{\partial w^*}{\partial r^*}\right) + \frac{\partial^2 w^*}{\partial z^{*2}}\right]$$

where

$$\varepsilon \equiv \frac{h}{b} \qquad \lambda \equiv h\sqrt{\frac{\Omega}{\nu}} \qquad (9)$$

By parameter $\lambda$ the Reynolds number is expressed as $\mathrm{Re} = \lambda^2$, and the Ekman number as $\mathrm{Ek} = 1/\lambda^2$. Note that $\lambda/h = \sqrt{\Omega/\nu}$ represents the boundary layer thickness. Note also that by omitting the terms in (8) which are multiplied by $\varepsilon^2$ one obtains boundary layer approximation equations. Introducing a volume flow rate coefficient defined as

$$C_w \equiv \frac{Q}{\nu b} \qquad (10)$$

the dimensionless form of volume flow (4) can be, by using (5), (6) and (9), expressed as

$$\int_{-1}^{1} u^*(1, z^*) dz^* = \frac{\varepsilon C_w}{2\pi \lambda^2} \qquad (11)$$

**4 Solution**

Equation (7) is solved by introducing the dimensionless stream function $\psi^*$ by which the dimensionless radial and tangential components of velocity are expressed as

$$u^* = \frac{1}{r^*} \frac{\partial \psi^*}{\partial z^*} \qquad w^* = -\frac{1}{r^*} \frac{\partial \psi^*}{\partial r^*} \qquad (12)$$

The stream function $\psi^*$ is sought to arrive at the following form of asymptotic series expansions

$$\begin{aligned}\psi^*(r^*, z^*; \varepsilon) &= \varepsilon^2 \sum_{n=0}^{\infty} \psi_n(z^*) \left(\frac{\varepsilon}{r^*}\right)^{2n} \\ &= \varepsilon^2 \psi_0(z^*) + \frac{\varepsilon^4 \psi_1(z^*)}{r^2} + \frac{\varepsilon^6 \psi_2(z^*)}{r^4} + \frac{\varepsilon^8 \psi_3(z^*)}{r^6} + \cdots\end{aligned} \qquad (13)$$

where $\psi_n(z^*)$ are new unknown functions. The form of $\psi^*$ was obtained by inspection and it ensure that $\varepsilon$ drops from resulting system of differential equation as it will be seen below. Once $\psi^*$ is known the asymptotic series expansion of radial and axial velocity components are, from (12)

$$u^*(r^*, z^*; \varepsilon) = \varepsilon \sum_{n=0}^{\infty} \psi'_n(z^*) \left(\frac{\varepsilon}{r^*}\right)^{2n+1}$$

$$= \frac{\varepsilon^2 \psi'_0(z^*)}{r^*} + \frac{\varepsilon^4 \psi'_1(z^*)}{r^{*3}} + \frac{\varepsilon^6 \psi'_2(z^*)}{r^{*5}} + \frac{\varepsilon^8 \psi'_3(z^*)}{r^{*7}} + \cdots \quad (14)$$

$$w^*(r^*, z^*; \varepsilon) = 2\varepsilon \sum_{n=1}^{\infty} n \psi_n(z^*) \left(\frac{\varepsilon}{r^*}\right)^{2n}$$

$$= \frac{2\varepsilon^3 \psi_1(z^*)}{r^2} + \frac{4\varepsilon^5 \psi_2(z^*)}{r^4} + \frac{6\varepsilon^7 \psi_3(z^*)}{r^6} + \cdots$$

where $(\ )' = d(\ )/dz^*$. Similarly, the series expansion for the tangential component of velocity and pressure are assumed to be

$$v^*(r^*, z^*; \varepsilon) = \varepsilon \sum_{n=0}^{\infty} v_n(z^*) \left(\frac{\varepsilon}{r^*}\right)^{2n+1}$$

$$= \frac{\varepsilon^2 v_0(z^*)}{r^*} + \frac{\varepsilon^4 v_1(z^*)}{r^{*3}} + \frac{\varepsilon^6 v_2(z^*)}{r^{*5}} + \frac{\varepsilon^8 v_3(z^*)}{r^{*7}} + \cdots \quad (15)$$

$$p^*(r^*, z^*; \varepsilon) = \varepsilon^2 \left[ p_0 \ln r^* + \sum_{n=1}^{\infty} p_n(z^*) \left(\frac{\varepsilon}{r^*}\right)^{2n} \right]$$

$$= \varepsilon^2 p_0 \ln r^* + \frac{\varepsilon^4 p_1(z^*)}{r^2} + \frac{\varepsilon^6 p_2(z^*)}{r^4} + \frac{\varepsilon^8 p_3(z^*)}{r^6} + \cdots$$

where $v_n(z^*)$ and $p_n(z^*)$ are new unknown functions. Substituting (14) and (15) into (8) and equating terms in equal powers of $r$ one obtains a system of successive linear differential equations

$$\psi'''_0 = \lambda^2(-2v_0 + p_0) \qquad v''_0 = 2\lambda^2 \psi'_0 \quad (16)$$

$$\psi'''_1 = -\lambda^2(2v_1 + 2p_1 + \psi'^2_0 + v_0^2) \qquad v''_1 = 2\lambda^2 \psi'_1 \qquad p'_1 = 0 \quad (17)$$

and for $n \geq 2$

$$\psi_n''' = -2\lambda^2 (v_n + n\, p_n) - 4n(n-1)\psi_{n-1}'$$
$$-\lambda^2 \sum_{k=0}^{n-1} \left[ (2k+1)\psi_k' \psi_{n-k-1}' - 2k\, \psi_k \psi_{n-k-1}'' + v_k\, v_{n-k-1} \right]$$
$$v_n'' = 2\lambda^2 \psi_n' - 4n(n-1)v_{n-1} + \lambda^2 \sum_{k=1}^{n-1} 2k \left( \psi_k v_{n-k-1}' - v_k \psi_{n-k-1}' \right) \quad (18)$$
$$p_n' = \frac{2}{\lambda^2}(n-1)\left[ 4(n-1)(n-2)\psi_{n-2} + \psi_{n-1}''' \right] - 4\sum_{k=1}^{n-2} k(n-2k-3)\psi_k \psi_{n-k-2}'$$

The boundary conditions for (16), (17) and (18) are from (3) and (11) by using (14) and (15), the following

$$\psi_0(1) - \psi_0(-1) = \frac{C_w}{2\pi\varepsilon\lambda^2}$$
$$\psi_0'(\pm 1) = v_0(\pm 1) = 0 \quad (19)$$

$$\psi_n(\pm 1) = \psi_n'(\pm 1) = v_n(\pm 1) = 0 \qquad (n = 1, 2, 3, ..) \quad (20)$$

The task is now to successively solve equations (16), (17) and (18) subject to boundary conditions (19) and (20) with $\varepsilon$, $\lambda$ and $C_w$ as input parameters.

*4. 1 First set of equations*

The first equation in (16) expresses

$$v_0 = -\frac{\psi_0'''}{2\lambda^2} + \frac{p_0}{2} \quad (21)$$

Substituting this into the second equation of (16) yields

$$\psi_0^{(5)} + 4\lambda^4 \psi_0' = 0 \quad (22)$$

The solution of this equation is

$$\psi_0(z^*) = C_{0,0} + C_{0,1} \operatorname{ch} \lambda z^* \sin \lambda z^* + C_{0,2} \operatorname{sh} \lambda z^* \cos \lambda z^* \\ + C_{0,3} \operatorname{ch} \lambda z^* \cos \lambda z^* + C_{0,4} \operatorname{sh} \lambda z^* \sin \lambda z^* \tag{23}$$

where $C_{0,0}, C_{0,1}, C_{0,2}, C_{0,3}, C_{0,4}$ are integration constants. From boundary conditions (19) one finds that $C_{0,3} = C_{0,4} = 0$ and

$$\begin{aligned} C_{0,1} &= \frac{C_w}{2\pi\varepsilon\lambda^2} \frac{\operatorname{sh}\lambda \sin\lambda - \operatorname{ch}\lambda \cos\lambda}{\operatorname{sh} 2\lambda - \sin 2\lambda} \\ C_{0,2} &= \frac{C_w}{2\pi\varepsilon\lambda^2} \frac{\operatorname{sh}\lambda \sin\lambda + \operatorname{ch}\lambda \cos\lambda}{\operatorname{sh} 2\lambda - \sin 2\lambda} \\ p_0 &= -\frac{C_w}{\pi\varepsilon\lambda^2} \frac{\operatorname{ch} 2\lambda + \cos 2\lambda}{\operatorname{sh} 2\lambda - \sin 2\lambda} \end{aligned} \tag{24}$$

The constant $C_{0,0}$ remains indeterminate but this does not affect the velocity components and pressure. Substituting (23) into (13), (14) and (21) yields the first approximation of components of velocity and pressure

$$\begin{aligned} u^* &= \frac{\lambda\varepsilon^2}{r^*} \left[ (C_{0,1} - C_{0,2}) \operatorname{sh}\lambda z^* \sin\lambda z^* + (C_{0,1} + C_{0,2}) \operatorname{ch}\lambda z^* \cos\lambda z^* \right] \\ v^* &= \frac{\varepsilon^2}{r^*} \left[ \frac{p_0}{2} + \lambda(C_{0,1} + C_{0,2}) \operatorname{sh}\lambda z^* \sin\lambda z^* - \lambda(C_{0,1} - C_{0,2}) \operatorname{ch}\lambda z^* \cos\lambda z^* \right] \\ p^* &= \varepsilon^2 p_0 \ln r^* \end{aligned} \tag{25}$$

These formulas are essentially the same as those of Breiter and Pohlhausen (1962).

*4.2 Second set of equations*

The solution of the second set of equations is obtained in a similar way. First, from the third equation in (17) it follows that $p_1 = \text{const}$ and from the first

$$v_1 = -\frac{\psi_1'''}{2\lambda^2} - p_1 - \frac{1}{2}\left(\psi_0'^2 + u_0^2\right) \tag{26}$$

Substituting (26) into the second equation of (17) and using (21) one finds

$$\psi_1^{(4)} + 4\lambda^4 \psi_1 = C_{1,0} - 2\lambda^2 \left(\psi_0'\psi_0'' + \lambda^2 p_0 \psi_0 - \psi_0 \psi_0'''\right) \tag{27}$$

The solution of (27) is obtained by the methods of variation of constants and is of the form

$$\begin{aligned}\psi_1(z) &= C_{1,0} + C_{1,1} \operatorname{ch} \lambda z \sin \lambda z + C_{1,2} \operatorname{sh} \lambda z \cos \lambda z \\ &\quad + C_{1,3} \operatorname{ch} \lambda z \cos \lambda z + C_{1,4} \operatorname{sh} \lambda z \sin \lambda z \\ &\quad + a_1 \lambda z \operatorname{ch} \lambda z \cos \lambda z + a_2 \lambda z \operatorname{sh} \lambda z \sin \lambda z + a_3 \left(\operatorname{sh} 2\lambda z - \sin 2\lambda z\right)\end{aligned} \tag{28}$$

where

$$a_0 = \frac{C_{1,0}}{4\lambda^4} \qquad a_1 = \frac{p_0}{8}\left(C_{0,1} + C_{0,2}\right)$$
$$a_2 = \frac{p_0}{8}\left(C_{0,1} - C_{0,2}\right) \qquad a_3 = -\frac{\lambda}{10}\left(C_{0,1}^2 + C_{0,2}^2\right) \tag{29}$$

and $C_{1,0}, C_{1,1}, C_{1,2}, C_{1,3}, C_{1,4}$ are integration constants. From (20) the boundary conditions are $\psi_1(\pm 1) = \psi_1'(\pm 1) = 0$, from which it follows that $C_{1,0} = C_{1,3} = C_{1,4} = 0$ and

$$\begin{aligned}2\Delta C_{1,1} &= -a_1\left(1 + \cos 2\lambda\right)\left(\operatorname{sh} 2\lambda - 2\lambda\right) - a_2\left(\sin 2\lambda + 2\lambda\right)\left(\operatorname{ch} 2\lambda - 1\right) \\ &\quad - 2a_3\left(\operatorname{sh} 3\lambda \cos \lambda + \operatorname{ch} \lambda \sin 3\lambda + \operatorname{ch} 3\lambda \sin \lambda - \operatorname{sh} \lambda \cos 3\lambda - 6\operatorname{sh} \lambda \cos \lambda\right) \\ 2\Delta C_{1,2} &= a_1\left(1 + \operatorname{ch} 2\lambda\right)\left(\sin 2\lambda - 2\lambda\right) - a_2\left(\operatorname{sh} 2\lambda + 2\lambda\right)\left(\cos 2\lambda - 1\right) \\ &\quad - 2a_3\left(\operatorname{sh} 3\lambda \cos \lambda + \operatorname{ch} \lambda \sin 3\lambda - \operatorname{ch} 3\lambda \sin \lambda + \operatorname{sh} \lambda \cos 3\lambda - 6\operatorname{ch} \lambda \sin \lambda\right)\end{aligned} \tag{30}$$

where $\Delta = \operatorname{sh} 2\lambda - \sin 2\lambda$. From (26) and condition $v_1(\pm 1) = 0$ one finds

$$p_1 = \lambda\left[\left(C_{1,2} - C_{1,1} - 3a_2\right)\operatorname{ch}\lambda + \lambda\left(a_1 - a_2\right)\operatorname{sh}\lambda\right]\cos\lambda$$
$$+ \lambda\left[\left(C_{1,1} + C_{1,2} + 3a_1\right)\operatorname{ch}\lambda + \lambda\left(a_2 + a_1\right)\operatorname{sh}\lambda\right]\sin\lambda - 4\lambda\left(\operatorname{ch}2\lambda + \cos 2\lambda\right) \quad (31)$$

This completes the solution of the second approximation. The explicit formulas for components of velocity and pressure are omitted because of their length and complexity.

*4.3 Third and fourth set of equations*

Only two additional sets of equations will be taken into account. From (18), when $n = 2$, the third set of equations is

$$\psi_2''' = -2\lambda^2 v_2 - 4\lambda^2 p_2 - 8\psi_1' - 2\lambda^2\left(2\psi_0'\psi_1' - \psi_0''\psi_1 + v_0 v_1\right)$$
$$v_2'' = 2\lambda^2\psi_2' - 8v_1 + 2\lambda^2\left(\psi_1 v_0' - \psi_0' v_1\right) \quad (32)$$
$$p_2' = \frac{2}{\lambda^2}\psi_1''$$

and for $n = 3$ the fourth set is

$$\psi_3''' = 4\lambda^2\psi_0''\psi_2 - 6\left(4 + \lambda^2\psi_0'\right)\psi_2' - 2\lambda^2 v_0 v_2 - 2\lambda^2 v_3 - 6\lambda^2 p_3$$
$$- \lambda^2\left(3\psi_1'^2 - 2\psi_1\psi_1'' + v_1^2\right)$$
$$v_3'' = 4\lambda^2 v_0'\psi_2 - 4\left(6 + \lambda^2\psi_0'\right)v_2 + 2\lambda^2\psi_3' - 2\lambda^2\left(\psi_1' v_1 - \psi_1 v_1'\right) \quad (33)$$
$$p_3' = \frac{4}{\lambda^2}\psi_2'' + 8\left(\frac{4}{\lambda^2} + \psi_0'\right)\psi_1$$

The boundary conditions for the above equations are from (20)

$$\psi_2(\pm 1) = \psi_2'(\pm 1) = v_2(\pm 1) = \psi_3(\pm 1) = \psi_3'(\pm 1) = v_3(\pm 1) = 0 \quad (34)$$

Since the structure of equations becomes more and more complex, the equations (32) and (33) are solved numerically as one set of six linear differential equations which together with (34) represent the linear boundary value problem.

**5 Example**

The obtained solution will now be compared with those obtained by Crespo del Arco et al (1996, Peyret 2002) for $C_w = 100$, $Ek = 2.24 \times 10^{-3}$, $R_m = \frac{b+a}{b-a} = 1.22$, $L = \frac{b-a}{2h} = 3.37$. These data givs $\lambda = 21.1289$ and $\varepsilon = 0.133665$. For the purpose of calculations a computer program was written. For the solution of the linear boundary value problem of the third and fourth set of equations the collocation boundary value solver *colnew* was used (Ascher et al 1995). All computations using the *colnew* subroutine were performed by setting tolerance to $10^{-8}$ for all variables. The positions of extremes of velocity components were calculated numerically by the function *fmin* (Forsythe et al 1977). Also it turns out that the obtained solutions of the problem are asymptotic series' which are divergent, so the question regarding how many terms are needed to calculate unknown functions accurately arises. Following Van Dyke (1975) the terms of divergent series' was at each radius summed up to the smallest.

The profiles of radial, tangential and axial velocity at $r^* = 0.54955$ (which correspond to Crespo del Arco's $\tilde{r} = 4.1$) are shown in Figures 2, 3 and 4. All computed velocity components were rescaled by the factor $\lambda^2/\varepsilon$ for purposes of comparison. It is seen from these figures that the profiles for radial and tangential velocity match those given in (Peyret 2002). In Figure 5 the distribution of pressure along disks at $z^* = 0$ is shown calculated using different numbers of terms in the solution. It is seen that for $r^* > 0.3$ the two, three and four term solution practically coincides, while for smaller values the three term solution is of use.

Table 1 compares results obtained by the present solution to those obtained by Hide's approximation and Crespo del Arco's numerical pseudo-spectral method for tangential velocity value at $z^* = 0$ and the position of maximal radial velocity and minimal tangential velocity. The minor discrepancy between solutions can be explained by the fact that boundary conditions for solutions at entrance and exit are different.

For additional confirmation of the solution, Table 2 gives the values and axial position of maximal radial velocity and values and position of minimal tangential velocity calculated using different numbers of terms in the series solution. It is seen from the table that computed maximal value of radial velocity when using one term and four term solutions differs by about 7%, while its position matches to two decimal places. The same observation also holds for minimal tangential velocity, but minimal value differs by about 9%. The radial velocity profile near maximal value for different numbers of terms used in calculation is also presented in Figure 6.

To estimate relative error the values of velocity components was calculated at a fixed axial position with different numbers of terms. The results of calculation with successive relative errors are shown in Table 3. It is seen that estimated relative error drops with the number of terms; thus relative error is at most 2%.

## 6. Concluding remarks

The present solution regarding the stationary flow of viscous fluid between two parallel co-rotating disks differs from known analytical solutions since no approximation is made in the governing continuity and the Navier-Stokes equations. However, the obtained solution for unknown velocity components and pressure in an asymptotic series form has a drawback since these series' are divergent meaning that all the problems associated with such series' are incorporated into the solution. Regardless of the utility of the contemporary numerical methods by which not only steady state but also turbulent flow can be studied, the present analytical solution has some methodological value and can also be used for comparative calculation when testing the accuracy of different numerical methods in fluid mechanics.

**References**

Acheson, D.J., 1990. Elementary Fluid Dynamics. Oxford University Press

Ames, W.F., 1965. Nonlinear Partial Differential Equations in Engineering. Academic Press.

Ascher, U.M., Mattheij, R.M.M., Russell, R.D., 1995. Numerical Solution of Boundary Value  Problems for Ordinary Differential Equations, SIAM,  Philadelphia

Bennetts, D.A., Jackson, W.D.N., 1974. Source-sink Flows in a Rotating Annulus - A Combined Laboratory and Numerical Study. J. Fluid Mech., **66**, 689-705.

Boyack, B.E., Rice, W., 1971. Integral Method for Flow Between Co-rotaning Disk, J. Basic Eng., Trans. ASME,  350-354

Boyd, K.E., Rice, W., 1968. Laminar Inward Flow of Incompressible Fluid Between Rotating Disks, with Full Peripheral Admission, J. App. Mech., Trans. ASME, pp. 229-237

Breiter, M.C., Pohlhausen, K., 1962. Laminar Flow Between Two Parallel Rotating Disks, ARL, USAF, Dayton, Ohio

Crespo del Arco, E., Maubert, P., Randriamampianina, A. and Bontoux, P. (1996) Spatio Temporal Behaviour in a Rotating Annulus with a Source-Sink Flow. J. Fluid Mech., **32**, 1-27.

Van Dyke, M., 1975. Perturbation Methods in Fluid Mechanics. The Parabolic Press, Stanford, California

Forsythe, G.E., Malcolm, M.A., C.B. Moler, 1977. Computer Methods For Mathematical Computations, Prentice-Hall, New Jersey.

Hide, R. 1968. On Source-Sink Flows Stratified in a Rotating Annulus. J. Fluid Mech., **32**, 737-764.

Landau, L.D., Lifshitz, E.M., 1987. Fluid Mechanics, 2nd edition, Butterworrth-Heineman

Owen, J.M., Pincombe, J.R., Rogers, R.H., 1985. Source-sink Flow Inside a Rotating Cylindrical Cavity. Journal of Fluid Mechanics 155, 233-265

Peyret. R., 2002. Spectral Methods for Incompressible Viscous Flow. Springer Verlag

Rajagopal, K.R., 1992. Flow of Viscoelastic Fluids Between Rotating Disks. Theoret. Comput. Fluid Dynamics 3, 185-206

Rice, W., 1963. An Analytical and Experimental Investigation of Multiple Disk Pumps and Compressors. Journal of Engineering for Power, Trans. ASME, 191-198

Savage, S.B., 1964. Laminar Radial Flow Between Parallel Plates, J. App. Mech., Trans. ASME, 594-596

**TABLES**

**Table 1.** Comparison of results obtained by present solution (A) to those obtained by Hide's approximation (B) and Crespo del Arco's numerical pseudo-spectral method (C)

| Tangential velocity $v(r^*,0)$ | | | Radial velocity $z_{max}$ | | | Tangential velocity $z_{min}$ | | |
|---|---|---|---|---|---|---|---|---|
| A | B | C | A | B | C | A | B | C |
| -668.6 | -569 | -639 | 0.9608 | 0.963 | 0.959 | 0.8815 | 0.889 | 0.876 |

**Table 2.** Values of and position of $u_{max}$ and $v_{min}$ for different numbers of terms *n* in series solution

| n | $z_{max}$ | $u_{max}$ | $z_{min}$ | $v_{min}$ |
|---|---|---|---|---|
| 1 | 0.9628 | 197.2 | 0.8885 | -652.9 |
| 2 | 0.9605 | 185.9 | 0.8832 | -690.4 |
| 3 | 0.9606 | 184.4 | 0.8818 | -707.8 |
| 4 | 0.9608 | 184.6 | 0.8815 | -719.6 |

**Table 3.** Values of $u_{max}$ at $z_{max}=0.9608$, $v(r^*,0)$ and $v_{min}$ at $z_{min}=0.8815$ for different numbers of terms ($r^*=0.54955$)

| n | $u_{max}$ | $|u_n/u_{n+1}-1|$ | $v(r^*,0)$ | $|v_n/v_{n+1}-1|$ | $v_{min}$ | $|v_n/v_{n+1}-1|$ |
|---|---|---|---|---|---|---|
| 1 | 196.9 | 0.06 | -611.9 | 0.05 | -652.1 | 0.06 |
| 2 | 185.9 | 0.01 | -642.5 | 0.02 | -690.4 | 0.02 |
| 3 | 184.5 | 0.00 | -658.0 | 0.02 | -707.8 | 0.01 |
| 4 | 184.6 | | -668.6 | | -719.6 | |

**FIGURES**

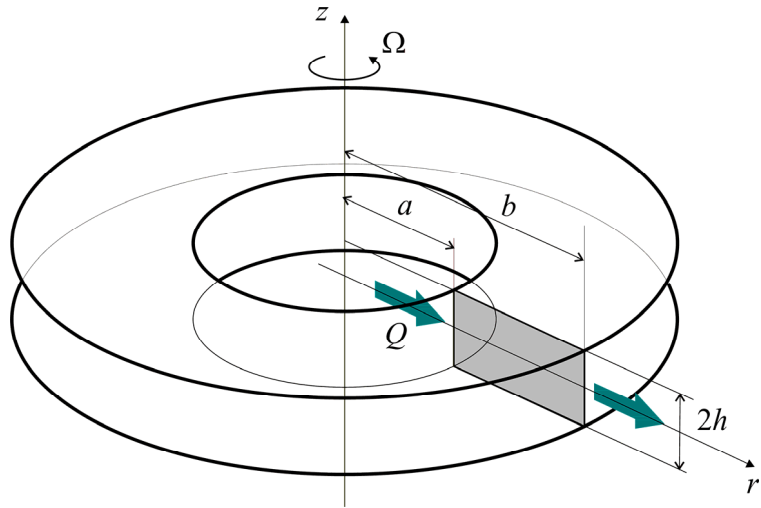

**Figure 1.** Geometry of the problem

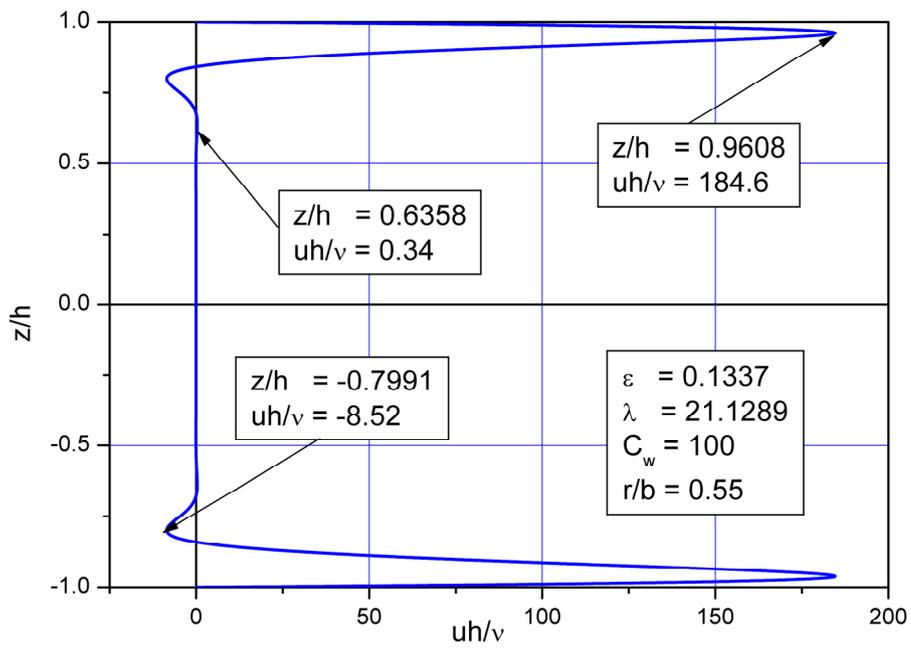

**Figure 2.** Radial velocity profile

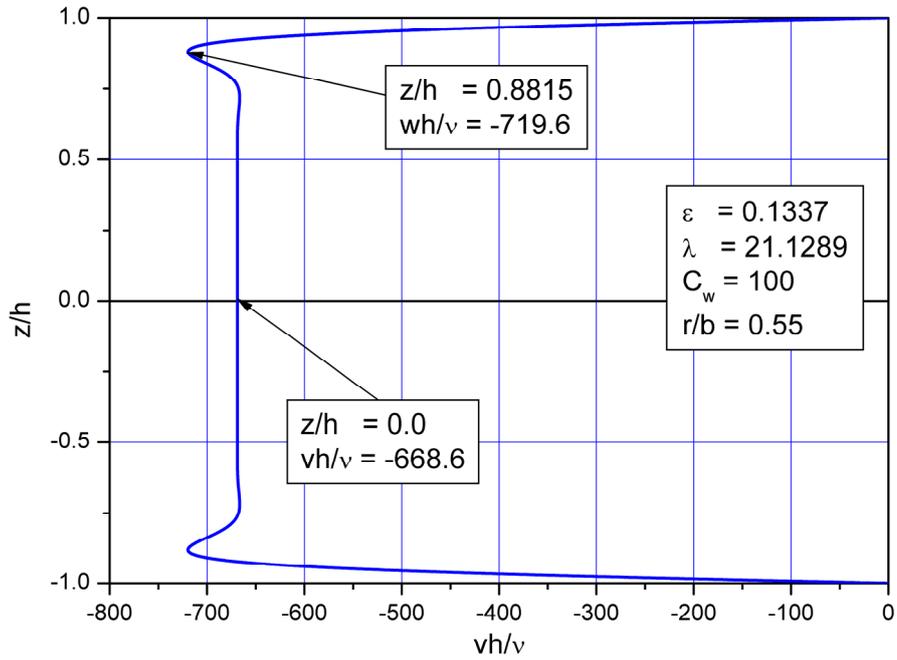

**Figure 3.** Tangential velocity profile

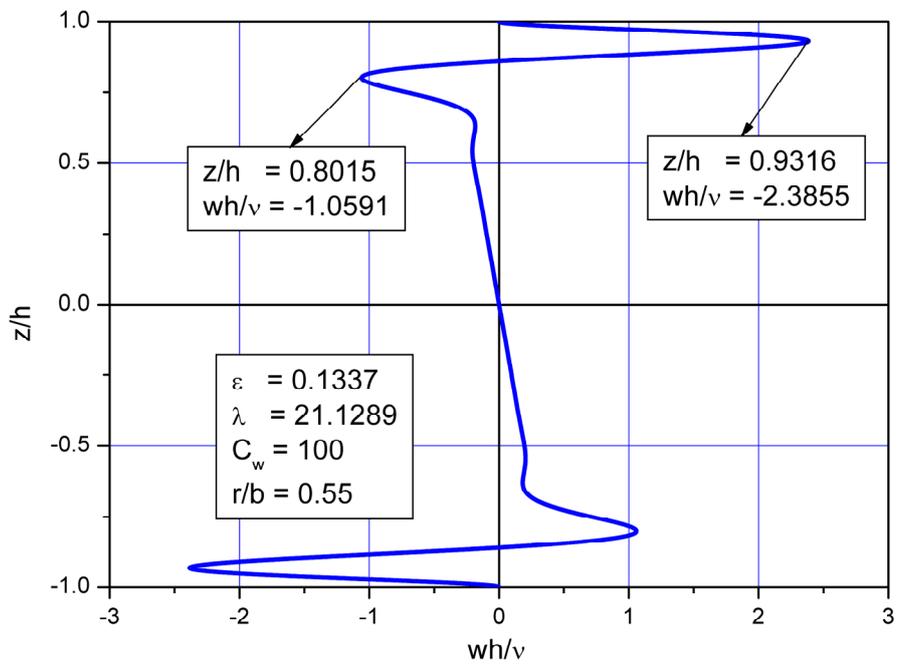

**Figure 4.** Axial velocity profile

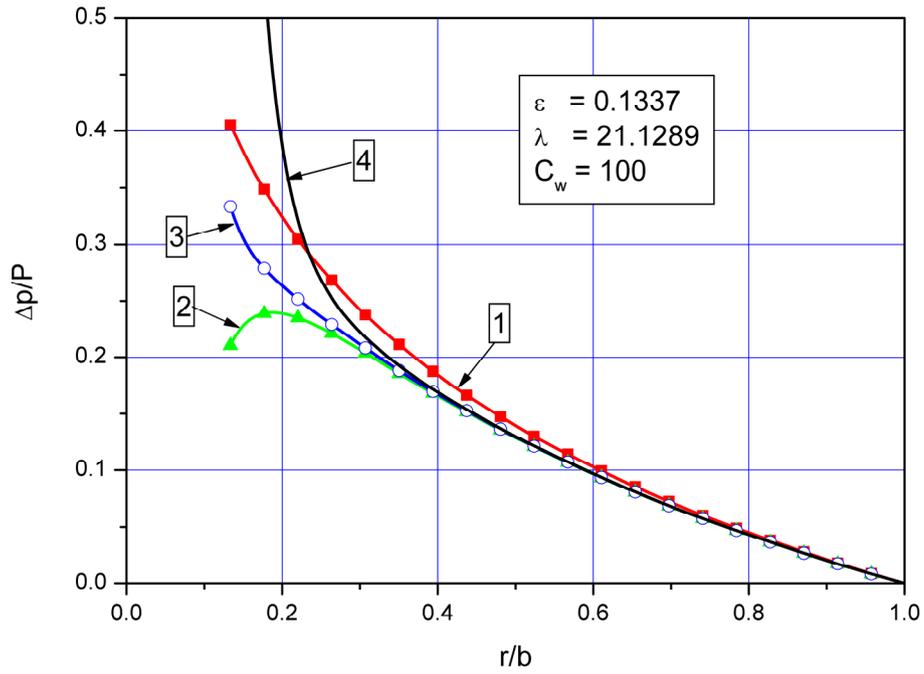

**Figure 5.** Reduced pressure distribution for various numbers of series terms used.

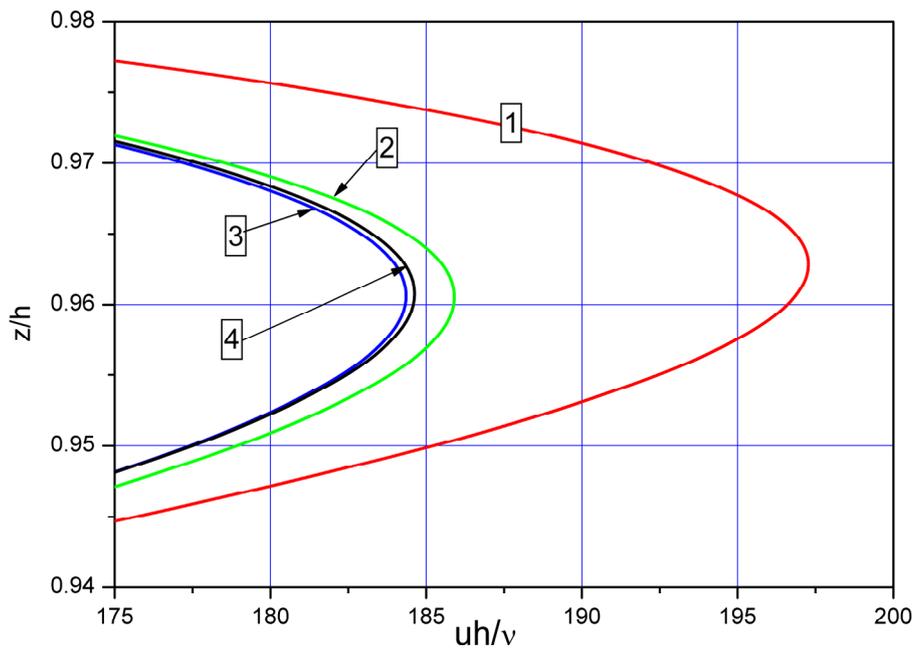

**Figure 6**. Detail of radial velocity profile near maximal value for different numbers of terms used in calculation for $r^* = 0.54955$.